\documentstyle[preprint,eqsecnum,aps,prb]{revtex}

\begin{document}
\title{Comment on ''the controlled charge ordering and evidence of the metallic
state in Pr$_{0.65}$Ca$_{0.35}$MnO$_{3}$ films''}
\author{W. Prellier\thanks{%
prellier@ismra.fr} and B.\ Raveau}
\address{Laboratoire CRISMAT, CNRS\ UMR 6508, Bd du\\
Mar\'{e}chal Juin, 14050 Caen Cedex, FRANCE.}
\date{\today}
\maketitle

\begin{abstract}
In a recent paper (2000 {\it J.\ Phys.: Condens. Matter} {\bf 12} L133) Lee 
{\it et al.} have studied the transport properties of Pr$_{0.65}$Ca$_{0.35}$%
MnO$_{3}$ thin films. They claimed that they are able to controlled the
charge-ordered (CO) state by the lattice strains. We propose herein another
alternative since another indexation of the orientation of the film can be
found leading to almost no distortion of the cell, as compared to the bulk
compound.
\end{abstract}

\newpage

In a recent paper dealings with the study of Pr$_{0.65}$Ca$_{0.35}$MnO$_{3}$
thin films, Lee {\it et al.}\cite{Lee} have shown that the ''appearance of
the charge-ordered (CO) state at 205\ K can be controlled via the lattice
strains accumulated during the film growth''

\medskip

Even if few papers have been published up to now on the particular compounds
showing the charge ordering state\cite{Rao1,Wil1,Wil2,Wagner}, \ the
manganite thin films have been extensively studied in the past years due to
their properties of colossal magnetoresistance, a huge decrease in
resistance when applying a magnetic field\cite{VonH,Mc,Venky,Jin}. These
authors present results of their X-Rays and resistivity measurements of Pr$%
_{0.65}$Ca$_{0.35}$MnO$_{3}$ thin films grown by the pulsed laser
deposition.. They obtain X-Rays diffraction (XRD) patterns highly resolved
peaks.\ Considering that these peaks correspond to the $00l$ reflections ($%
002$ and $004$ respectively) of the Pr$_{1-x}$Ca$_{x}$MnO$_{3}$ type
structure, they demonstrate that the as-grown thin film of the manganite is
tetragonal with $a=b=5.432$\AA\ and $c=7.74$\AA , i.e. fundamentally
different form the bulk material with the same composition Pr$_{0.65}$Ca$%
_{0.35}$MnO$_{3}$ which is orthorhombic with $a=5.42$\AA , $b=5.46$\AA\ and $%
c=7.67$\AA \cite{PCMOb}. Unfortunately, for the determination of the
structure, they also use weak reflections of the perovskite ($103$, $113$
and $312$) which appear on they XRD\ pattern, but which belong to a
secondary phase with a different orientation and possibly with a different
composition (note, that the cationic composition of the film has not been
checked and that a small change in composition can drastically modify the
parameters of the cell and therefore the properties).Nevertheless, the
authors conclude that lattice strains are at the origin of the wired
behavior of this thin film which exhibits, according to them, a metallic
like;e behavior with a incredibly high resistivity of 10$^{5}$ $\Omega .cm$
below 120K ! After, a brief summary on the charge-ordering, we will show
that another conclusion can be proposed leading to no distortion of the cell
as compared to the bulk material.

\bigskip Hole-doped manganites, with the general formula RE$_{1-x}$A$_{x}$MnO%
$_{3}$ (RE=rare earth, A=alkaline earth) exhibit a rich phase diagram as a
function of the doping concentration $x$. As the average size of the cations
at the lanthanide site is reduced, the tilt of the MnO$_{6}$ octahedra is
induced which favors the localization and ordering of the Mn$^{3+}$/Mn$^{4+}$
cations\cite{Rao}. Moreover, below \ a certain temperature ($T_{CO}$), the
metallic state becomes unstable and the material goes to an insulating
state. Such a charge-ordering transition is associated with
antiferromagnetic, insulating properties and large lattice distortions.
Another amazing feature is that the application of a magnetic field results
in the melting of the CO\ state (i.e. the material becomes ferromagnetic
metallic). This phenomenon has been observed in Pr$_{1-x}$Ca$_{x}$MnO$_{3}$
(with $0.35<x<0.65$) \cite{PCMOa}. Thus, the composition Pr$_{0.65}$Ca$%
_{0.35}$MnO$_{3}$ studied in the described paper shows in bulk, a CO state
at low temperature.

We propose another alternative concerning \ the crystallographic nature of
this Pr$_{0.65}$Ca$_{0.35}$MnO$_{3}$ thin film. It has been shown previously 
\cite{Berny,AMH}, using electron diffraction, that many manganite thin films
with a $Pbnm$ space group (as Pr$_{0.65}$Ca$_{0.35}$MnO$_{3}$) , are often
[110]-oriented when grown on LaAlO$_{3}$ (as used by Lee {\it et al.}).
Thus, it can instead be proposed that these peaks on the XRD\ pattern do not
correspond to the $00l$ reflections but to the $hk0$ reflections (i.e. $110$
and $220$ respectively). As a consequence, the d2200 and d004 observed
values and deduced from the datas obtained by the authors ($%
d_{220}^{obs}\cong 1.93$\AA\ and $d_{004}^{obs}\cong 1.92$\AA\ for the
as-grown and $d_{220}^{obs}\cong 1.92$\AA\ $d_{004}^{obs}\cong 1.96$\AA\ for
the annealed film) are perfectly compatible with those calculated from the
parameters of the cell of the corresponding bulk compound ($%
d_{220}^{cal}\cong d_{004}^{cal}\cong 1.92$\AA \cite{PCMOb}). In others
words the XRD\ pattern of the Pr$_{0.65}$Ca$_{0.35}$MnO$_{3}$ thin film
suggests that the film is identical to the bulk, and this does not bring any
proof about the accumulation of the lattice strains during the film growth,
contrary to the statement of the authors.

\medskip

In conclusion, the assumption upon the orientation of the film of Pr$_{0.65}$%
Ca$_{0.35}$MnO$_{3}$ made by these authors is purely speculative. This
demonstrates that the thin films must be correctly characterized from the
crystal structure viewpoint before going to physical measurements, using
electron diffraction to reconstruct the reciprocal space. For this reason,
all the physics which is developed in this paper might be uncorrected.

\end{document}